\documentclass[aps,prl,twocolumn,a4paper,showpacs,superscriptaddress,
               amsmath]{revtex4}

   \usepackage{graphicx}

\newcommand{\expect}[1]{\mathinner{\langle #1\rangle}}
\newcommand{\Expect}[1]{\left< #1 \right>}

\newcommand{\comm}[2]{\left[#1,#2\right]}
\newcommand{\anticomm}[2]{\left\{#1,#2\right\}}

\newcommand{\dd}{\mathrm{d}}

\def\clap#1{\hbox to 0pt{\hss#1\hss}}

\def\mathrlap{\mathpalette\mathrlapinternal}

\def\mathrlapinternal#1#2{\rlap{$\mathsurround=0pt#1{#2}$}}

\newcommand{\Obog}{O_{\mathrm{Bog}}}
\newcommand{\OFT}{O_{\mathrm{FT}}}
\newcommand{\Gammavac}{\Gamma_{\mathrm{vac}}}

\begin{document}

\title{Engineering correlation and entanglement dynamics in spin
  systems}

\date{January 9, 2007}

\pacs{03.67.-a, 03.67.Mn}

\author{T. S. Cubitt}
\affiliation{Max Planck Institut f\"ur Quantenoptik,
  Hans--Kopfermann Str.\ 1, D-85748 Garching, Germany}
\affiliation{Department of Mathematics, University of Bristol,
  University Walk, Bristol BS8 1TW, UK}
\author{J.I. Cirac}
\affiliation{Max Planck Institut f\"ur Quantenoptik,
  Hans--Kopfermann Str.\ 1, D-85748 Garching, Germany}

\begin{abstract}
  We show that the correlation and entanglement dynamics of spin
  systems can be understood in terms of propagation of spin
  waves. This gives a simple, physical explanation of the behaviour
  seen in a number of recent works, in which a localised, low-energy
  excitation is created and allowed to evolve. But it also extends to
  the scenario of translationally invariant systems in states far from
  equilibrium, which require less local control to prepare. Spin-wave
  evolution is completely determined by the system's dispersion
  relation, and the latter typically depends on a small number of
  external, physical parameters. Therefore, this new insight into
  correlation dynamics opens up the possibility not only of predicting
  but also of \emph{controlling} the propagation velocity and
  dispersion rate, by manipulating these parameters. We demonstrate
  this analytically in a simple, example system.
\end{abstract}

\maketitle

Correlations play a predominant role in the study of spin systems. On
the one hand, they characterize different phases of matter, and thus
can help reveal the mechanisms underlying phase transitions. On the
other, they are directly related to the entanglement between different
spins, which can be exploited by applications in the field of quantum
information processing. Whereas so far, much of the work on
correlations has focused on the static properties of equilibrium
systems, an increasing interest in the corresponding dynamical
properties has developed over the last few years. The reason is two
fold. Firstly, new experimental setups, such as atoms in optical
lattices, have reached an unprecedented level of control, allowing
physical parameters to be changed during the experiments. Thus
theoretical descriptions of the time-dependent properties of such
systems have become important. Secondly, it has been recognized that
the way entanglement is created and how it propagates are important
fundamental questions in quantum information theory. In particular,
the answers may influence the design of quantum repeaters and
networks, whose goal is to establish as much entanglement as possible
between different nodes in the shortest possible time.

The time evolution of correlation functions in spin systems has been
studied recently in various scenarios, mainly from a condensed matter
physics perspective. In Refs.~\cite{IR2000,AOP+2004}, two-point
correlations were studied numerically, whereas in Ref.~\cite{CC2006}
their evolution in a critical model was studied analytically using
conformal field theoretic methods. In all cases, correlations were
seen to propagate at a finite speed. In Ref.~\cite{BHF2006}, a proof
was given that correlations \emph{necessarily} propagate at a finite
speed. On the other hand, information and entanglement propagation in
spin systems has mostly been studied from a quantum information
perspective \cite{OL2004,Bose2003,CDEL2004,CVC2005}.

In contrast with previous work, we will consider to what extent it is
possible to \emph{control} the propagation speed and dispersion of the
correlations in a translationally invariant system, by tuning only
simple, global, physical parameters. This may be relvant for the
optimal creation of entanglement in spin systems, as well as
contributing to a better understanding of how correlations are created
in dynamical processes, something which can be tested experimentally
in present setups. We will show that, even with this severely limited
control over the system, the correlation speed can be engineered
whilst simultaneously keeping dispersion to a minimum, so that
correlations can be concentrated between particular spins. Indeed, by
manipulating system parameters during the evolution, the speed can be
adjusted at will, even to the extent of reducing it to zero, allowing
correlations to be frozen at a desired location.

It is instructive to first consider the entanglement and correlation
propagation described in the references given above from a new
perspective. In many of those works, correlation propagation can be
understood as follows. The spin system is initially prepared in its
ground state. A localised, low-energy excitation is then created
(e.g.\ by flipping one spin), and allowed to evolve. Since the
low-energy excitations take the form of spin waves, the correlation
and entanglement dynamics can be understood as nothing other than
propagation of spin waves. This is completely determined by the
dispersion relation (given by the system's spectrum). The form of the
dispersion relation will typically depend on external, physical
parameters of the system (e.g.\ the strength of an external magnetic
field). Thus already in these setups, we can manipulate the external
parameters to control the dispersion relation, and hence control the
propagation of correlations. For example, changing the gradient of the
dispersion relation will change the propagation speed.

However, the ground state will typically be highly-correlated and
difficult to prepare, and with the level of local control required to
create the local excitation and break the translational symmetry, more
sophisticated quantum-repeater setups are possible. Also, it is not
clear that the correlations will remain localised; they are likely to
disperse rapidly as they propagate.

Therefore, we will extend the idea to systems prepared in
translationally invariant, easily created, uncorrelated initial
states. For example, the fully polarised state with all spins aligned
can be prepared by applying a large, external magnetic field. As the
initial state will be far from the ground state, it will contain many
excitations. The correlation dynamics is then the result of the
propagation and interference of a large number of spin waves at many
different frequencies. Nonetheless, we will show analytically that, at
least for some simple models, the system can be engineered so that
correlations propagate in well-defined, localised wave packets, with
little dispersion. The external parameters can then be used to control
the propagation of these correlation packets.

In the following, we will consider a specific model which, despite its
simplicity, is sufficiently rich to display most of the features we
are interested in. The model is simple enough to envisage implementing
it experimentally, for instance using atoms in optical lattices or
trapped ions. The XY--model for a chain of spin--$\frac{1}{2}$
particles is described by the Hamiltonian $H_{XY} =
-\tfrac{1}{4}\sum_{l}((1+\gamma)\sigma^x_l\sigma_{l+1} +
(1-\gamma)\sigma^y_l\sigma^y_{l+1} + 2\lambda\sigma^z_l)$, where the
$\sigma$'s are the usual Pauli operators and the sum is over spin
indices. The parameter $\lambda$ can be interpreted as the strength of
a global, external magnetic field, whereas $\gamma$ controls the
anisotropy of the interactions.

This Hamiltonian can be brought into diagonal form by the well-known
procedure \cite{Sachdev} of applying Jordan-Wigner, Fourier and
Bogoliubov transformations, giving $H_{XY} =
-\tfrac{i}{4}\sum_k\varepsilon_k(\gamma^x_k\gamma^p_k -
\gamma^p_k\gamma^x_k)$ with spectrum $\varepsilon_k = ((\cos(2\pi
k/N)-\lambda)^2 + \gamma^2\sin^2(2\pi k/N))^{1/2}$. The
$\gamma^{x,p}_k$ are Majorana operators, related to the more usual
Jordan-Wigner fermionic annihilation operator $\gamma_k$ by
$\gamma^x_k=\gamma_k^\dagger + \gamma_k$ and
$\gamma^p_k=(\gamma_k^\dagger-\gamma_k)/i$, and obey canonical
anti-commutation relations
$\anticomm{\gamma^a_k}{\gamma^b_l}=2\delta_{k,l}\delta_{a,b}$.

Ultimately, we are interested in ``connected'' spin--spin correlation
functions, for example the ZZ correlation function $C_{ZZ}(n,m) =
\expect{\sigma^z_n\sigma^z_m} -
\expect{\sigma^z_n}\expect{\sigma^z_m}$, in which the ``classical''
part of the correlation is subtracted. These are related to the
localisable entanglement $L(n,m)$ (the maximum average entanglement
between two spins $n$ and $m$ that can be extracted by local
measurements on all the others \cite{VPC04}): the natural figure of
merit for quantum repeaters. In particular, for spin--$\tfrac{1}{2}$
systems, $L(n,m) \geq C(n,m)$ for any connected spin--spin correlation
function $C$ \cite{PVMC05}. However, we will start by considering the
simpler, albeit less well-motivated, string correlation functions such
as $S_{XX}(i,j) =
\expect{\sigma^x_i(\prod_{i<k<j}\sigma^z_k)\sigma^x_j}$ (important for
revealing ``hidden order'' in certain models \cite{AKLT88}). Their
behaviour will give insight into the more important spin--spin
correlations, and we will use similar techniques to calculate both.

Assume the spin chain is initially in some completely separable,
uncorrelated state, such as the state with all spins down. The
interactions are then switched on and, as this initial state is not an
eigenstate of the Hamiltonian (unless $\lambda\to\infty$), the state
evolves in time. The initial state is the vacuum of the Majorana
operators $x_l\,(p_l) = \prod_j\sigma^z_j\sigma^{x(y)}_l$ obtained
after applying just the Jordan-Wigner transformation, and is
completely determined by its two-point correlation functions. In other
words, the vacuum is a fermionic Gaussian state, and can be
represented by its covariance matrix $\Gamma_{m,n} =
\tfrac{1}{2}\Expect{\comm{r_m}{r_n}}$ where $r_{2l-1}=x_l$ and
$r_{2l}=p_l$.

From the Heisenberg evolution equations, it is simple to show that any
evolution governed by a quadratic Hamiltonian corresponds to an
orthogonal transformation of the covariance matrix. It is also clear
that, as the Fourier and Bogoliubov transformations are canonical
(anti-commutation-relation-preserving) transformations of the Majorana
operators, they similarly leave Gaussian states Gaussian, and they too
can be expressed as orthogonal transformations. Thus the time-evolved
state of the system is given by a series of orthogonal transformations
of the fermionic vacuum:
\begin{equation}\label{eq:gamma}
  \Gamma(t) = \mathcal{O}\Gammavac\mathcal{O}^T,
  \quad\mathcal{O} = \OFT^T\,\Obog^T\,O(t).
\end{equation}
This is a block-Toeplitz matrix, composed of $2\times 2$ blocks $G_x$
at distance $x$ from the main diagonal. In the thermodynamic limit
$N\to\infty$ with $\frac{2\pi k}{N}\to\phi$ and
$\varepsilon_k\to\varepsilon(\phi)\equiv\varepsilon$,
\begin{gather*}
  G_x = \int_{-\pi}^\pi\dd\phi
        \begin{pmatrix}g_0&g_1\\g_{-1}&g_0\end{pmatrix},\quad
  g_0 = i S \sin(\phi x)\sin\bigl(2\varepsilon t\bigr)\\
  \begin{split}
    g_{\pm 1}
      = 2 C S \sin(\phi x)&\sin^2\bigl(\varepsilon t\bigr)
        \pm \cos(\phi x)
         \Bigl(C^2 + S^2\cos\bigl(2\varepsilon t\bigr)\Bigr)
  \end{split}
\end{gather*}
where $C = (\cos(\phi) - \lambda)/\varepsilon(\phi)$, $S =
\gamma\sin(\phi)/\varepsilon(\phi)$, and $x = m-n$. We can now
calculate certain string correlations, which are given directly by
elements of the covariance matrix. For example,
$\bigl\langle\sigma^x_n(\prod_{n<i<m}\sigma^z_i)
\sigma^y_m\bigr\rangle = \tfrac{1}{i}\Gamma_{2n-1,2m-1} = \sum_{s=\pm
  1} s\int_{-\pi}^{\pi}\dd\phi S \cos(\phi x + 2s\varepsilon t)/2$.

Although the evolution of the string correlations is produced by the
collective dynamics of a large number of excitations, this expression
has a simple, physical interpretation: it is the equation for two wave
packets with envelope $S/2$ propagating in opposite directions along
the chain, according to a dispersion relation given by the system's
spectrum $\varepsilon(\phi)$. This wave-packet interpretation allows
us to make quantitative predictions as to how the dynamics will be
affected if the system parameters $\gamma$ and $\lambda$ are
modified. Specifically, modifying the parameters will change the
dispersion relation, changing the group velocity of the correlation
packets, as well as the rate at which they disperse. (The wave-packet
envelopes also depend on the system parameters, so the relevant region
of the dispersion relation may also change.) Thus by varying only
global physical parameters, we can control the speed at which
correlations propagate.

Does this hold true for the more interesting spin--spin correlations?
We will show analytically that they have a similar wave-packet
description, although in terms of multiple packets propagating
simultaneously. This will allow us to predict the behaviour of the
spin--spin correlation dynamics for different values of the system
parameters. In particular, we will show that the correlations can be
made to propagate in well-defined packets whose speed can be
engineered by tuning the system parameters. Moreover, the propagation
speed can be controlled as the system is evolving, so we can speed up
or slow down the packets, even to the extent of reducing the speed to
zero. We confirm our predictions by numerically evaluating the
analytic expressions.

Let us now calculate the spin--spin connected correlation function
using the covariance matrix derived above. We have $\sigma^z_n =
x_np_n$, so the ZZ connected correlation function is given by
$C_{ZZ}(x) = \expect{x_np_m}\expect{p_nx_m} -
\expect{x_nx_m}\expect{p_np_m}$, where we have used Wick's theorem to
expand the expectation value of the product of four Majorana operators
into a sum of expectation values of pairs \cite{Wick50,CPC06}. The
latter are given by covariance matrix elements, resulting in the
following analytic expression for the correlation function:
\begin{equation}
  \label{eq:ZZ}
  \begin{split}
    &C_{zz}(x,t)^2 =\\
      &\biggl(\int_{-\pi}^\pi\dd \phi\,
         \frac{S}{2}\sum_{s=\pm 1}\bigl(
           s\cos(\phi x - 2s\varepsilon t)
         \bigr)
       \biggr)^2\\
      &+\biggl(\int_{-\pi}^\pi\dd \phi\,
         CS \bigl(\sin(\phi x) - \frac{1}{2}\sum_{s=\pm 1}
                  \sin(\phi x + 2s\varepsilon t)\bigr)
       \biggr)^2\\
      &-\biggl(
        \int_{-\pi}^\pi\dd \phi\,
        \bigl(C^2\cos(\phi x) + \frac{S^2}{2}\sum_{s=\pm 1}
              \cos(\phi x + 2s\varepsilon t)\bigr)
      \biggr)^2.
  \end{split}
  \raisetag{11.5em}
\end{equation}

\begin{figure}[!htbp]
  \includegraphics{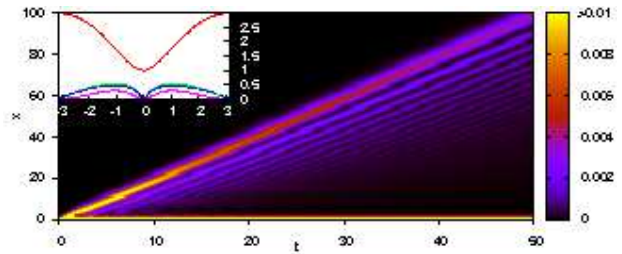}
  \caption{For $\gamma=1.1$, $\lambda=2$, all the wave-packet
    envelopes from Eq.~\eqref{eq:ZZ} (non-red curves, inset) are
    similar in form, centred around a nearly linear region of the
    dispersion relation with gradient $\approx 2$ (red curve,
    inset). Thus the correlations $C_{zz}(x,t)$ (indicated by the
    shading, main plot) propagate in well-defined packets at a speed
    given by the gradient.}
  \label{fig:slow-packets}
\end{figure}

Although more complicated than the string correlations, this
expression also describes wave packets evolving according to the same
dispersion relation $\varepsilon(\phi)$, albeit multiple packets with
different envelopes propagating and interfering simultaneously (three
in each direction). In many parameter regimes, broad (in
frequency-space) wave packets and a highly non-linear dispersion
relation will cause the correlations to rapidly disperse and
disappear. However, we can find regimes in which the wave packets are
located in nearly linear regions of the dispersion relation, and
maintain their shape as they propagate. For example, at $\gamma=1.1$
and $\lambda=2$, all three wave packets of Eq.~\eqref{eq:ZZ} are
nearly identical, and reside in an almost-linear region of the
dispersion relation with gradient roughly equal to 2, as shown in
Fig.~\ref{fig:slow-packets} (inset). The spin--spin correlation
dynamics will therefore involve well-defined correlation packets
propagating at a speed $\dd x/\dd t=2$, dispersing only slowly as they
propagate. Fig.~\ref{fig:slow-packets} shows the result of numerically
evaluating Eq.~\eqref{eq:ZZ}, which clearly confirms the predictions.

\begin{figure}[!htbp]
  \includegraphics{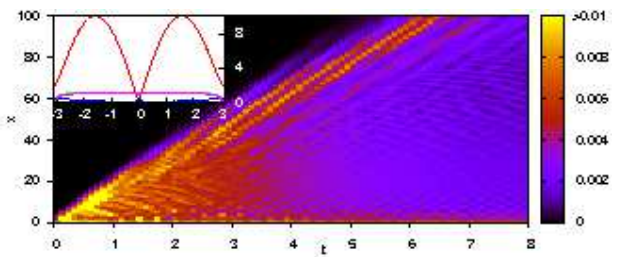}
  \caption{For $\gamma=10$, $\lambda=0.9$, the wave-packet envelopes
    are spread over the entire frequency range (non-red curves,
    inset). However, the dispersion relation (red curve, inset) is
    almost linear for wavenumbers not too near $\pi$, with gradients
    $\pm 18$. As the envelopes are symmetric about $\pi$, most of the
    correlations $C_{zz}(x,t)$ (shading, main plot) still propagate at
    a well-defined speed $\approx 18$, faster than in
    Fig.~\ref{fig:slow-packets} (note scale), though as expected they
    also show more dispersion.}
  \label{fig:fast-packets}
\end{figure}

We can engineer a different correlation speed by changing the
parameters. For instance, for $\gamma=10$ and $\lambda=0.9$ we predict
a higher propagation speed $\dd x/\dd t \approx 18$, although at the
expense of increased dispersion. The numerical results of
Fig.~\ref{fig:fast-packets} show precisely this behaviour.

\begin{figure}[!htbp]
  \includegraphics{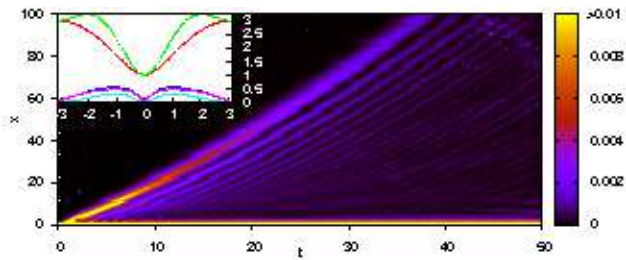}
  \caption{Starting from $\gamma=1.1$, $\lambda=2$ as in
    Fig.~\ref{fig:slow-packets}, $\gamma$ and $\lambda$ are smoothly
    changed to move from the red to the green dispersion relation
    (inset), increasing the correlation speed.}
  \label{fig:speed-up}
\end{figure}

An even more interesting possibility is controlling the correlation
packets as they propagate. If the system parameters are changed
continuously in time, the XY-Hamiltonian becomes time-dependent, and
the orthogonal evolution operator $O(t)$ in Eq.~\eqref{eq:gamma} is
given by a time-ordered exponential $O(t) = T[e^{\int_0^t\dd t'
  A(t')}] \equiv \lim_{h\to 0}\prod_{n=1}^{\lfloor t/h \rfloor}
e^{A(nh)}$. ($A$ is a time-dependent, anti-symmetric matrix determined
by the Hamiltonian.) In general, the time-ordering is essential. But
if the system parameters change slowly in time, dropping it will give
a good approximation to the evolution operator. The state at time $t$
is then just given by evolution under the time-average (up to $t$) of
the Hamiltonian. If we remain in a parameter regime for which the
relevant region of the dispersion relation is nearly linear, adjusting
the parameters changes the gradient without significantly affecting
its curvature or the form of the wave packets. Thus, to good
approximation, slowly adjusting the parameters should control the
speed of the wave packets as they propagate, allowing us to speed them
up and slow them down. Numerically evaluating the time-ordered
exponential shows this is indeed possible (Fig.~\ref{fig:speed-up}).

Clearly it would be useful to be able to stop the correlations once
they reach a desired location. One way would be to simply switch off
the interactions. But strictly speaking this would require more
control than is provided by the two parameters defined in the
Hamiltonian (there is no value of $\gamma$ for which all interaction
terms vanish), and may be difficult in physical implementations. If
the spin model were realised in a solid-state system, for example,
switching off the interactions would likely involve fabricating an
entirely new system. In any case, we will show that switching off the
interactions is not necessary in order to freeze correlations at a
specific location.

\begin{figure}[!htbp]
  \includegraphics{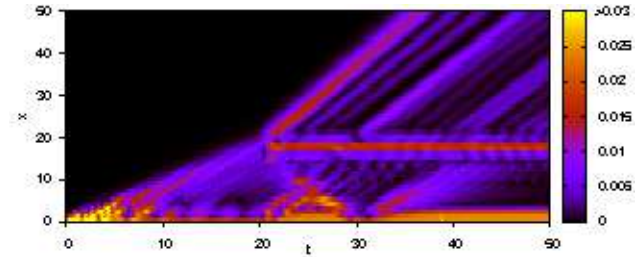}
  \caption{The system is initially allowed to evolve with
    $\gamma_0=0.9$ $\lambda_0=0.5$, then ``quenched'' at time $t_1=20$
    to $\gamma_1=0.1$, $\lambda_1=10$. Some of the correlations are
    frozen at the separation ($x\approx 20$) they reached at
    $t_1$. Others propagate according to the new dispersion relation,
    or are ``reflected''.}
  \label{fig:quench}
\end{figure}

Instead of changing the parameters continuously, we now consider
changing them abruptly. The time-evolved covariance matrix in this
scenario can be calculated analytically by the same methods as used
above. Suppose the initial system parameters $\gamma_0$ and
$\lambda_0$ are suddenly changed to $\gamma_1$ and $\lambda_1$ at time
$t_1$. The spin--spin correlations will initially evolve according to
Eq.~\eqref{eq:ZZ}, as before. After time $t_1$, the evolution becomes
more complicated. The analogue of Eq.~\eqref{eq:ZZ} separates into a
sum of wave packets evolving in four different ways: those that
initially evolve according to $\varepsilon_0$ and subsequently (after
$t_1$) evolve according to $\varepsilon_1$, those that subsequently
evolve according to $-\varepsilon_1$, those that only start evolving
at $t_1$, and those that that undergo no further evolution after $t_1$
(Fig.~\ref{fig:quench}).  For $t>t_1$, the terms whose evolution is
``frozen'' at time $t_1$ are given by
\begin{align*}
  &C_z^{t_1}(x,t)^2 =\\
  &\left(\frac{1}{2}\int_{\mathrlap{-\pi}}^{\mathrlap{\pi}}\dd \phi
       S_0S_1(C_0S_1-C_1S_0)
       \sum_{s=\pm 1} \sin(\phi x + 2s t_1\varepsilon_0)
     \right)^2\\
  &-\left(\frac{1}{2}\int_{\mathrlap{-\pi}}^{\mathrlap{\pi}}\dd \phi
       S_0C_1(C_0S_1-C_1S_0)
       \sum_{s=\pm 1} \cos(\phi x + 2s t_1\varepsilon_0)
     \right)^{\mathrlap{2}}.
\end{align*}
Since $t$ does not appear on the right hand side, this expression
clearly describes wave packets that propagate until time $t_1$ and
then stop. Using these, we can move correlations to the desired
location, then ``quench'' the system by abruptly changing the
parameters, freezing the correlations at that location, as shown in
Fig.~\ref{fig:quench}.

We have shown that entanglement and correlation propagation in many
spin-model setups can be understood in terms of propagation of spin
waves, and have introduced the idea of controlling the dynamics via
their dispersion relation, by manipulating external parameters of the
system. Although this is in principle possible for almost any spin
system, preparing a single-excitation initial state would require
control over individual spins. Therefore, we have analysed in detail
the more complex case of systems prepared in uncorrelated,
translationally invariant initial states, which typically contain many
excitations. We have shown for an example model that the dynamics can
be described by a small number of correlation wave packets, and that
the control afforded by a few external, physical parameters is
sufficient to allow detailed control over the propagation of
correlations.

\bibliography{XY-Edyn}

\end{document}